\documentclass[preprint,aps,prl,superscriptaddress,showpacs]{revtex4}

\usepackage{graphicx}
\begin{document}


\title{Evidence for Narrow $S$=+1 Baryon Resonance in Photo-production from 
Neutron\\}


\author{T.~Nakano}
  \affiliation{Research Center for Nuclear Physics, Osaka University, Ibaraki, Osaka 567-0047, Japan}
\author{D.S.~Ahn}
  \affiliation{Department of Physics, Pusan National University, Busan 609-735, Korea}
\author{J.K.~Ahn}
  \affiliation{Department of Physics, Pusan National University, Busan 609-735, Korea}
\author{H.~Akimune}
  \affiliation{Department of Physics, Konan University, Kobe, Hyogo 658-8501, Japan}
\author{Y.~Asano}
  \affiliation{Synchrotron Radiation Research Center, Japan Atomic Energy Research Institute, Mikazuki, Hyogo 679-5198, Japan}
  \affiliation{Advanced Science Research Center, Japan Atomic Energy Research Institute, Tokai, Ibaraki 319-1195, Japan}
\author{W.C.~Chang}
  \affiliation{Institute of Physics, Academia Sinica, Taipei 11529, Taiwan}
\author{S.~Dat\'e}
  \affiliation{Japan Synchrotron Radiation Research Institute, Mikazuki, Hyogo 679-5198, Japan}
\author{H.~Ejiri}
  \affiliation{Japan Synchrotron Radiation Research Institute, Mikazuki, Hyogo 679-5198, Japan}
  \affiliation{Research Center for Nuclear Physics, Osaka University, Ibaraki, Osaka 567-0047, Japan}
\author{H.~Fujimura}
  \affiliation{School of Physics, Seoul National University, Seoul, 151-747, Korea}
\author{M.~Fujiwara}
  \affiliation{Research Center for Nuclear Physics, Osaka University, Ibaraki, Osaka 567-0047, Japan}
  \affiliation{Advanced Science Research Center, Japan Atomic Energy Research Institute, Tokai, Ibaraki 319-1195, Japan}
\author{K.~Hicks}
  \affiliation{Department of Physics and Astronomy, Ohio University, Athens, Ohio 45701}
\author{T.~Hotta}
  \affiliation{Research Center for Nuclear Physics, Osaka University, Ibaraki, Osaka 567-0047, Japan}
\author{K.~Imai}
  \affiliation{Department of Physics, Kyoto University, Kyoto 606-8502, Japan} 
\author{T.~Ishikawa}
  \affiliation{Laboratory of Nuclear Science, Tohoku University, Sendai, Miyagi 982-0826, Japan}
\author{T.~Iwata}
  \affiliation{Department of Physics, Yamagata University, Yamagata 990-8560, Japan}
\author{H.~Kawai}
  \affiliation{Department of Physics, Chiba University, Chiba 263-8522, Japan}
\author{Z.Y.~Kim}
  \affiliation{School of Physics, Seoul National University, Seoul, 151-747, Korea}
\author{K.~Kino}
  \affiliation{Research Center for Nuclear Physics, Osaka University, Ibaraki, Osaka 567-0047, Japan}
\author{H.~Kohri}
  \affiliation{Research Center for Nuclear Physics, Osaka University, Ibaraki, Osaka 567-0047, Japan}
\author{N.~Kumagai}
  \affiliation{Japan Synchrotron Radiation Research Institute, Mikazuki, Hyogo 679-5198, Japan}
\author{S.~Makino}
  \affiliation{Wakayama Medical University, Wakayama, Wakayama 641-8509, Japan}
\author{T.~Matsumura}
  \affiliation{Research Center for Nuclear Physics, Osaka University, Ibaraki, Osaka 567-0047, Japan}
  \affiliation{Advanced Science Research Center, Japan Atomic Energy Research Institute, Tokai, Ibaraki 319-1195, Japan}
\author{N.~Matsuoka}
  \affiliation{Research Center for Nuclear Physics, Osaka University, Ibaraki, Osaka 567-0047, Japan}
\author{T.~Mibe}
  \affiliation{Research Center for Nuclear Physics, Osaka University, Ibaraki, Osaka 567-0047, Japan}
  \affiliation{Advanced Science Research Center, Japan Atomic Energy Research Institute, Tokai, Ibaraki 319-1195, Japan}
\author{K.~Miwa}
  \affiliation{Department of Physics, Kyoto University, Kyoto 606-8502, Japan} 
\author{M.~Miyabe}
  \affiliation{Department of Physics, Kyoto University, Kyoto 606-8502, Japan} 
\author{Y.~Miyachi}
  \altaffiliation[Present address, ]{Department of Physics, Tokyo Institute of Technology, Tokyo 152-8551, Japan}
  \affiliation{Department of Physics and Astrophysics, Nagoya University, Nagoya, Aichi 464-8602, Japan}
\author{M.~Morita}
  \affiliation{Research Center for Nuclear Physics, Osaka University, Ibaraki, Osaka 567-0047, Japan}
\author{N.~Muramatsu}
  \affiliation{Advanced Science Research Center, Japan Atomic Energy Research Institute, Tokai, Ibaraki 319-1195, Japan}
\author{M.~Niiyama}
  \affiliation{Department of Physics, Kyoto University, Kyoto 606-8502, Japan} 
\author{M.~Nomachi}
  \affiliation{Department of Physics, Osaka University, Toyonaka, Osaka 560-0043, Japan}
\author{Y.~Ohashi}
  \affiliation{Japan Synchrotron Radiation Research Institute, Mikazuki, Hyogo 679-5198, Japan}
\author{T.~Ooba}
  \affiliation{Department of Physics, Chiba University, Chiba 263-8522, Japan}
\author{H.~Ohkuma}
  \affiliation{Japan Synchrotron Radiation Research Institute, Mikazuki, Hyogo 679-5198, Japan}
\author{D.S.~Oshuev}
  \affiliation{Institute of Physics, Academia Sinica, Taipei 11529, Taiwan}
\author{C.~Rangacharyulu}
  \affiliation{Department of Physics and Engineering Physics, University of Saskatchewan, Saskatoon, Saskatchewan, Canada, S7N 5E2} 
\author{A.~Sakaguchi}
  \affiliation{Department of Physics, Osaka University, Toyonaka, Osaka 560-0043, Japan}
\author{T.~Sasaki}
  \affiliation{Department of Physics, Kyoto University, Kyoto 606-8502, Japan} 
\author{P.M.~Shagin}
  \altaffiliation[Present address, ]{School of Physics and Astronomy, University of Minnesota, Minneapolis, Minnesota 55455}
  \affiliation{Research Center for Nuclear Physics, Osaka University, Ibaraki, Osaka 567-0047, Japan}
\author{Y.~Shiino}
  \affiliation{Department of Physics, Chiba University, Chiba 263-8522, Japan}
\author{H.~Shimizu}
  \affiliation{Laboratory of Nuclear Science, Tohoku University, Sendai, Miyagi 982-0826, Japan}
\author{Y.~Sugaya}
  \affiliation{Department of Physics, Osaka University, Toyonaka, Osaka 560-0043, Japan}
\author{M.~Sumihama}
  \affiliation{Department of Physics, Osaka University, Toyonaka, Osaka 560-0043, Japan}
  \affiliation{Advanced Science Research Center, Japan Atomic Energy Research Institute, Tokai, Ibaraki 319-1195, Japan}
\author{H.~Toyokawa}
  \affiliation{Japan Synchrotron Radiation Research Institute, Mikazuki, Hyogo 679-5198, Japan}
\author{A.~Wakai}
  \altaffiliation[Present address, ]{Akita Industry Promotion Foundation, Akita 010-8572, Japan}
  \affiliation{Center for Integrated Research in Science and Engineering, Nagoya University, Nagoya, Aichi 464-8603, Japan}
\author{C.W.~Wang}
  \affiliation{Institute of Physics, Academia Sinica, Taipei 11529, Taiwan}
\author{S.C.~Wang}
  \altaffiliation[Present address, ]{Institute of Statistical Science, Academia Sinica, Nankang, 115 Taipei, Taiwan}
  \affiliation{Institute of Physics, Academia Sinica, Taipei 11529, Taiwan}
\author{K.~Yonehara}
  \altaffiliation[Present address, ]{Department of Physics, University of Michigan, Ann Arbor, Michigan 48109-1120}
  \affiliation{Department of Physics, Konan University, Kobe, Hyogo 658-8501, Japan}
\author{T.~Yorita}
  \affiliation{Japan Synchrotron Radiation Research Institute, Mikazuki, Hyogo 679-5198, Japan}
\author{M.~Yoshimura}
  \affiliation{Institute for Protein Research, Osaka University, Suita, Osaka 565-0871, Japan}
\author{M.~Yosoi}
  \affiliation{Department of Physics, Kyoto University, Kyoto 606-8502, Japan} 
\author{R.G.T.~Zegers}
  \affiliation{Research Center for Nuclear Physics, Osaka University, Ibaraki, Osaka 567-0047, Japan}


\date{\today}

\begin{abstract}

The $\gamma n \rightarrow K^+  K^-  n$ reaction on $^{12}$C has
been studied by measuring both $K^+$ and $K^-$ at forward angles. A
sharp baryon resonance peak was observed at $1.54 \pm 0.01$ GeV/$c^2$
with a width smaller than 25 MeV/$c^2$ and a Gaussian significance
of 4.6 $\sigma$.  The strangeness quantum number ($S$) of the baryon
resonance is +1. It can be interpreted as a molecular meson-baryon
resonance or alternatively as an exotic 5-quark state ($uudd\bar{s}$)
that decays into a $K^+$ and a neutron.  The resonance is
consistent with the lowest member of an anti-decuplet of baryons
predicted by the chiral soliton model.

\end{abstract}


\maketitle

The search for baryon resonances with the strangeness quantum number
$S$=+1, that cannot be formed by three quarks, has a long and
interesting history.  In fact, the summary of the $S$=+1 baryon
resonance searches has been dropped from the Particle Data Group (PDG)
listings although the possible exotic resonances were noted in the
1986 baryon listings \cite{pdg86}.  Most of the previous searches were
made using the partial wave analyses of kaon-nucleon ($KN$)
scatterings~\cite{arndt}.  These searches resulted in two
possibilities, the isoscalar $Z_0$(1780) and $Z_0$(1865), for which
the evidence of the existence was reviewed to be poor by PDG.

The present work was motivated in part by the recent work by Diakonov,
Petrov and Polyakov \cite{dpp97} who studied anti-decuplet baryons
using the chiral soliton model. The mass splittings of the established
octet and decuplet were reproduced within accuracy of 1 \% in this
model, and those of the new anti-decuplet were also estimated using
the nucleon sigma term \cite{gasser} and the current quark-mass
ratios. The anti-decuplet was anchored to the mass of the
P$_{11}$(1710) nucleon resonance, giving the $Z^+$ (spin $1/2$,
isospin 0 and $S$=+1) a mass of $\sim 1530$ MeV/$c^2$ and a total
width of less than 15 MeV/$c^2$. The $S$=+1 baryon resonances in this
mass region have not been searched for in the $KN$ scattering
experiments in the past because momenta of kaons were too
high as pointed out in Refs.~\cite{dpp97,pol00}.

The concept of a molecular meson-baryon bound state has been proposed
by Refs.~\cite{lutz,oset,kaiser} in conjunction with the well-known
$\Lambda (1405)$ particle. The mass spectrum of the $\Lambda (1405)$
can be dynamically generated~\cite{lutz,oset} suggesting that this
``particle'' can be described as a molecular meson-baryon bound state
with a quark configuration $uuds\bar{u}$. However, the validity of
this assumption is not solid and should be checked by measuring the
$\Lambda(1405)$ decay~\cite{lutz,oset}, of which experimental data are
scarce. Moreover, the same quantum numbers for the $\Lambda(1405)$ can
be achieved with a quark configuration $uds$. This ambiguity is not
present in the case of the proposed $Z^+$ resonance with a quark
configuration $uudd\bar{s}$.  In this letter we report the
experimental evidence for a narrow resonance with $S = +1$ which can
be interpreted as the predicted exotic $Z^+$ state.

The experiment was carried out at the Laser-Electron Photon facility
at SPring-8 (LEPS)~\cite{nakano,nakano2,zegers}.  Photons were
produced by Compton back-scattering of laser photons from 8 GeV
electrons in the SPring-8 storage ring. Using a 351-nm Ar laser,
photons with a maximum energy of 2.4 GeV were produced. Electrons that
were participants in the back-scattering process were
momentum-analyzed by a bending magnet of the SPring-8 storage ring,
and detected by a tagging counter inside the ring to get the photon
energy with a resolution ($\sigma$) of 15 MeV. Only photons with
energies above 1.5 GeV were tagged.  The typical photon flux was $\sim
10^6$/s.

A 0.5-cm thick plastic scintillator (SC) which was composed of
hydrogen and carbon nuclei (C:H $\approx $ 1:1) was used as a target
in the present experiment (see Ref.~\cite{nakano}). The SC was located
9.5 cm downstream from the 5-cm thick liquid-hydrogen (LH$_2$) target
used for studying the photo-production of $\phi$-mesons.  In fact, the
two experiments were carried out simultaneously. Since this paper
concentrates on the study of events generated from neutrons in carbon
nuclei at the SC, a comparison between events from the LH$_2$ and the
SC selected under the same conditions, with only a change in the
software condition on the reconstructed vertex position ($vtz$) along
the beam axis, provides a good tool to distinguish contributions from
protons and neutrons.

A silicon-strip vertex detector (SSD) and 3 drift chambers were used
to track charged particles through a dipole magnet with a field
strength of 0.7 Tesla.  The SSD consists of single-sided silicon-strip
detectors (vertical and horizontal planes) with the strip pitch of 120
$\mu$m. The first drift chamber located before the magnet consists of
6 wire planes (3 vertical planes, 2 planes at $+ 45^\circ$, and 1
plane at $- 45^\circ$), and the other two drift chambers after the
magnet consist of 5 planes (2 vertical planes, 2 planes at $+
30^\circ$, and 1 plane at $- 30^\circ$).  A time-of-flight (TOF)
scintillator array was positioned 3 m behind the dipole magnet.
Electron-positron pairs produced at very forward angles were blocked
by lead bars which were set horizontally along the median plane inside
the magnet gap.  Electrons and positrons that escaped from the
blocker, and pions with a momentum higher than $\sim 0.6$ GeV/$c$ were
vetoed on-line by an aerogel ${\rm \check{C}}$erenkov counter located
downstream of the SC.

The angular coverage of the spectrometer was about $\pm$0.4 rad and
$\pm$0.2 rad in the horizontal and vertical directions,
respectively. The momentum resolution ($\sigma$) for
1-GeV/$c$ particles was 6 MeV/$c$. The timing resolution ($\sigma$) of
the TOF was 150 psec for a typical flight length of 4 m from the
target to the TOF. Particle identification was made within 3$\sigma$
of the momentum-dependent mass resolution, which was about 30
MeV/$c^2$ for a 1-GeV/$c$ kaon.

The design of the LEPS detector is optimized for measuring
$\phi$-mesons produced near the threshold energy at forward angles by
detecting the $K^+K^-$ pair from the $\phi$ decay. These measurements
will be reported in a separate article. Here we discuss the detection
of $K^+K^-$ pairs generated at the SC.  From the total set of $4.3
\times 10^{7}$ events measured in the LEPS detector, $8.0 \times
10^{3}$ events with a $K^+ K^-$ pair were selected.  As shown in
Fig.~1(a), a cut on the $vtz$ cleanly selected events that originate
from a reaction at the SC, which accounted for about half of the $K^+
K^-$-pair events.

To reduce contributions from non-resonant $K^+K^-$ productions for
which the phase space increases quadratically with the photon energy
from the production threshold, events with the photon energy above
2.35 GeV were rejected. About $3.2 \times 10^{3}$ events remained
after this cut. The missing mass $MM_{\gamma K^+K^-}$ of the ${\rm
N}(\gamma,K^+K^-){\rm X}$ reaction was calculated by assuming that the
target nucleon (proton or neutron) has the mean nucleon mass of 0.9389
GeV/$c^2$ ($M_N$) and zero momentum. Subsequently, events with
$0.90 < MM_{\gamma K^+K^-} < 0.98$ GeV/$c^2$ were selected.  A total of
$1.8 \times 10^{3}$ events survived after this cut. Most of the
remaining events ($\sim 85$ \%) were due to the photo-production of the
$\phi$ meson.  They were eliminated by removing the events with the
invariant $K^+K^-$ mass from 1.00 GeV/$c^{2}$ to 1.04 GeV/$c^{2}$ for
the $\phi$ (Fig.~1(b)).

In order to eliminate photo-nuclear reactions of $\gamma p \rightarrow
K^+ K^- p$ on protons in $^{12}$C and $^{1}$H at the SC, the recoiled
protons were detected by the SSD. The direction and momentum of the
nucleon in the final state was calculated from the $K^+$ and $K^-$
momenta, and such events in which the recoiled nucleon was out of the
SSD acceptance were rejected. Events were rejected if the momentum of
the nucleon was smaller than 0.35 GeV/$c$ since the calculated
direction had a large uncertainty in this case.  Finally, we rejected
108 events for which the hit position in the SSD agreed with the
expected hit position within 45 mm in the vertical or horizontal
direction. The cut points correspond to about $\pm$ 2$\sigma$
resolution for events that are affected by the Fermi motion. A total
of 109 events satisfied all the selection criteria. We call this set
of events the ``signal sample''.

In case of reactions on nucleons in nuclei, the Fermi motion has to be
taken into account to obtain appropriate missing-mass spectra.  To
evaluate this effect, we studied the $\gamma n \rightarrow K^+
\Sigma^- \rightarrow K^+ \pi^- n$ sequential process as an example,
where the $K^+$ and $\pi^-$ were detected. The missing masses,
$MM_{\gamma K^+}$ and $MM_{\gamma K^+\pi^-}$, were obtained for the
${\rm N}(\gamma,K^+){\rm X}$ and ${\rm N}(\gamma,K^+\pi^-){\rm N}$
channels by assuming that the nucleon in $^{12}$C is at rest with 
the mass equal to $M_N$.  Both
the missing masses are smeared out due to the Fermi motion of 
nucleons in $^{12}$C. However, since the nucleons in the two channels 
are identical, the two missing masses have a strong
correlation as shown in Fig.~2(a). Accordingly, the missing mass
corrected for the Fermi motion, $MM_{\gamma K^+}^c$, is deduced as
\begin{eqnarray}
MM_{\gamma K^+}^c = MM_{\gamma K^+} - MM_{\gamma K^+\pi^-} + M_N.
\end{eqnarray}
The correction in Eq.~1 compensates spreads not only due to the Fermi
motion but also due to the experimental resolutions and the binding
energy of the nucleon in the initial state, although the Fermi motion
effect is the major contribution here.  Note that this correction is
not a good approximation for events with $MM_{\gamma K^+\pi^-}$ far
from the nucleon rest mass.  The missing mass distributions before and
after the correction are compared in Fig.~2(b). Only after the
correction, the $\Lambda$ (from $\gamma p \rightarrow K^+ \Lambda
\rightarrow K^+ \pi^- p$) and the $\Sigma^{-}$ peaks are separated.
The correction is good in case of a decay with a small $Q$ value,
where the velocity of the hyperon is nearly the same as that of the
decaying nucleon. This is seen in the small width of the $\Lambda$ in
the missing mass spectrum in Fig.~2(b), where about half of the
contributions are due to reactions on protons in $^{12}$C. On the
other hand, the large width of the mass spectrum for the $\Sigma^-$ is
due to the imperfection of the correction caused by a large $Q$
value. The spectrum with a measured width ($\sigma$) of 18 MeV/$c^2$
is well reproduced by a Monte Carlo simulation using the impulse
approximation for the reaction from a neutron whose momentum
distribution is generated according to a harmonic oscillator
potential. The Monte Carlo simulation shows that the width is
dominated by incomplete cancellation of the Fermi motion effect.
Contributions of the binding energy in $^{12}$C to the peak position
and the width in the corrected spectrum are likely smaller than 10
MeV/$c^2$.

The missing mass $MM_{\gamma K^\pm}^c$ for $K^+K^-$ events is
corrected for the Fermi motion in the similar procedure. The corrected
missing mass is given by
\begin{eqnarray}
MM_{\gamma K^\pm}^c = MM_{\gamma K^\pm} - MM_{\gamma K^+K^-} + M_N.
\end{eqnarray}
In Fig.~3(a), the corrected $K^+$ missing-mass distribution for the 109
events that satisfy all the selection conditions is compared with that
for the 108 events for which a coincident proton hit was detected in
the SSD.  In the latter case, a clear peak due to the $\gamma+p
\rightarrow K^{+} \Lambda(1520) \rightarrow K^+ K^- p$ reaction is
observed. The $\Lambda(1520)$ peak does not exist in the case that the
proton-rejection cut in the SSD is applied as shown in the signal
sample.  This indicates that the signal sample is dominated by events
produced by reactions on neutrons.

Fig.~3(b) shows the corrected $K^-$ missing mass distribution of the
signal sample. A prominent peak at 1.54 GeV/$c^{2}$ is found. It
contains 36 events in the peak region $1.51 \leq MM_{\gamma K^-}^c <
1.57$ GeV/$c^{2}$. The broad background centered at $\sim 1.6$
GeV/$c^{2}$ is most likely due to non-resonant $K^+K^-$ production
because the events in the bump do not show any noticeable structure in
the $K^+$ missing mass nor in the invariant $K^+K^-$ mass spectra and
the beam-energy dependence of the production rate reflects the phase
space expansion with the energy. To estimate the background level due
to the non-resonant $K^+K^-$ production in the peak region of $1.51
\leq MM_{\gamma K^-}^c < 1.57$ GeV/$c^{2}$, the missing mass
distribution of the signal sample in the region above 1.59 GeV/$c^{2}$
was fitted by a distribution of events from the LH$_2$. For the event
selection, we applied the same conditions as for the signal selection
except that the $vtz$ window was shifted to the LH$_2$ position and a
proton hit in the SSD was allowed.  The $\Lambda(1520)$ contribution
in the LH$_2$ sample was eliminated by rejecting events with $1.51
\leq MM_{\gamma K^+}^c < 1.53$ GeV/$c^{2}$ because the $\Lambda(1520)$
contribution was removed from the signal sample by the requirement of
no-proton hit in the SSD.  The best fit with a $\chi^2$ of 7.2 for 8
degrees of freedom is obtained with a scale factor of 0.20 as shown
with a dotted histogram in Fig.~3(b).  Note that the acceptance of the
kaon pairs from the LH$_2$ is about 10 \% smaller than that from the
SC. However, the difference in the two acceptances does not depend on
$MM_{\gamma K^-}^c$ very much, and its effect on the background
estimation is negligible.

The background level in the peak region is estimated to be 17.0 $\pm$
2.2 $\pm$ $1.8$, where the first uncertainty is the error in the
fitting in the region above 1.59 GeV/$c^{2}$ and the second is a
statistical uncertainty in the peak region. The combined uncertainty
of the background level is $\pm$ 2.8. The estimated number of the
events above the background level is 19.0 $\pm$ 2.8, which corresponds
to a Gaussian significance of 4.6 $^{+1.2}_{-1.0}$ $\sigma$
($19.0/\sqrt{17.0} = 4.6$).  After subtracting the background from the
signal sample, the spectrum in the region of $1.47 \leq MM_{\gamma
K^-}^c < 1.61$ GeV/$c^{2}$ was compared with Monte Carlo simulations
assuming a Breit-Wigner function for a resonance distribution. The
best fit to the spectrum gives the mass of the resonance to be 1.54
$\pm$ 0.01 GeV/$c^2$, where the uncertainty is statistical only. The
systematic error was estimated to be 5 MeV/$c^2$ from that the
observed $\Sigma^-$ peak position was 5 MeV/$c^2$ smaller than the PDG
value of 1.197 GeV/$c^2$.  However, there could be a small shift due
to nuclear medium effects since the observed resonance was produced in
$^{12}$C. This point will be cleared up by experiments using a
deuterium target or in other reactions~\cite{dpp97,pol00}.  The width
$\Gamma$ of the resonance cannot be determined by the fitting since
the zero width gives the minimum $\chi^2$ of 1.6 for 4 degrees of
freedom. The upper limit for the width was determined to be 25
MeV/$c^2$ with a confidence level of 90 \%.

To make sure that the observed peak is not due to particle
misidentification of (one of) the kaons, the mass cut was tightened
from 3$\sigma$ to 2$\sigma$. One event in the peak was lost. This is
consistent with the statistical expectation due to a tightened cut.
Therefore, particle misidentification as a source of the peak is
unlikely. It was also confirmed that the peak was not generated by
intentionally selecting pions and by assuming the kaon mass in the
missing mass calculations. The $MM_{\gamma K^-}^c$ distributions for
events both at the $\phi$ resonance and its tails were checked.  The
shapes of the distributions were similar to that for the LH$_2$
background sample.  Thus, the observed peak is not due to the effect
of the tails of the $\phi$ resonance. No statistically significant
excess was found in the peak region for the events from the SC for
which a proton hit was found in the SSD. This further confirms that
the observed peak is due to the $\gamma n$ reactions.

In conclusion, we have performed a search for an $S = +1$ baryon
resonance in the $K^-$ missing mass spectrum of the $\gamma n
\rightarrow K^+ K^- n$ reaction on $^{12}$C, using a newly built
photon-beam facility at the SPring-8.  A sharp baryon resonance peak
has been found at $1.54 \pm 0.01$ GeV/$c^2$ in the $K^-$ missing
mass spectrum that is corrected for the Fermi motion.  The Gaussian
significance of the peak is 4.6 $\sigma$ and the width is estimated to
be smaller than 25 MeV/$c^2$. This strongly indicates the existence of an
$S = +1$ resonance which may be attributed to the molecular
meson-baryon resonance or alternatively as an exotic 5-quark baryon
proposed as the $Z^+$.

\acknowledgments

The authors gratefully acknowledge the dedicated efforts of the staff
of the SPring-8 for providing a good quality beam. We thank
Dr. T. Sato (Osaka Univ.), Dr. A. Hosaka (RCNP), and Dr. A. Titov
(JINR) for helpful discussions. This research was supported in part by
the Ministry of Education, Science, Sports and Culture of Japan, by
the National Science Council of Republic of China (Taiwan), and by
KOSEF of Republic of Korea.

\begin{figure}[H]
\scalebox{0.9}{\includegraphics{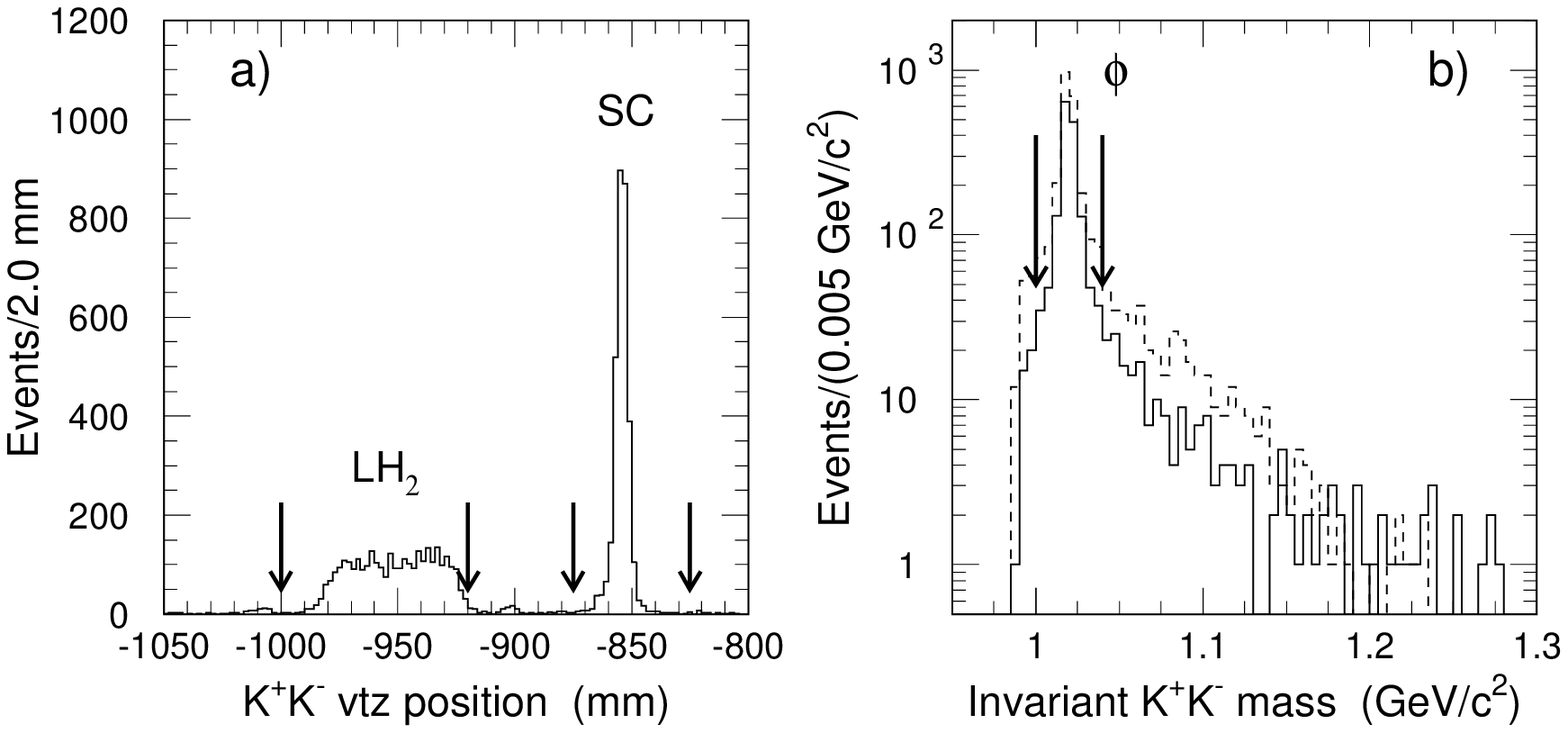}}
\caption{a) Vertex position ($vtz$) for $K^+K^-$ events along the photon-beam
direction. Cut points to select the SC events or the LH$_2$ events are
indicated by the arrows.  b) Invariant $K^-K^+$ mass distributions for
the SC events (solid histogram) and the LH$_2$ events (dashed
histogram). Cut points to exclude $\phi$ contributions are indicated
by the arrows.}
\end{figure}

\begin{figure}[H]
\scalebox{0.9}{\includegraphics{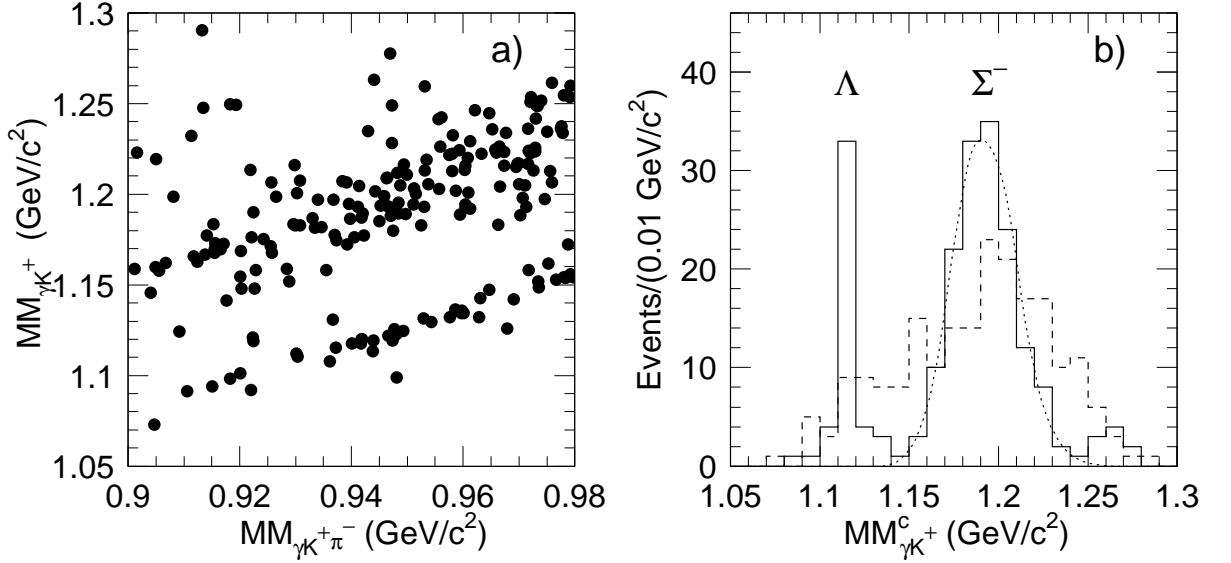}}
\caption{a) A scatter plot of $MM_{\gamma K^+}$ vs. $MM_{\gamma
K^+\pi^-}$ for $K^+\pi^-$ photo-productions at the SC. b) The missing
mass, $MM_{\gamma K^+}^c$, spectra for the $K^+\pi^-$ events from the
SC (solid histogram) and for Monte Carlo events for the $\gamma n
\rightarrow K^+ \Sigma^-$ channel (dotted curve) calculated via
Eq.~1. The dashed histogram shows the missing mass spectrum without
the Fermi-motion correction, $MM_{\gamma K^+}$ the projection of the
events in a) on to the vertical axis. }
\end{figure}

\begin{figure}[H]
\scalebox{0.9}{\includegraphics{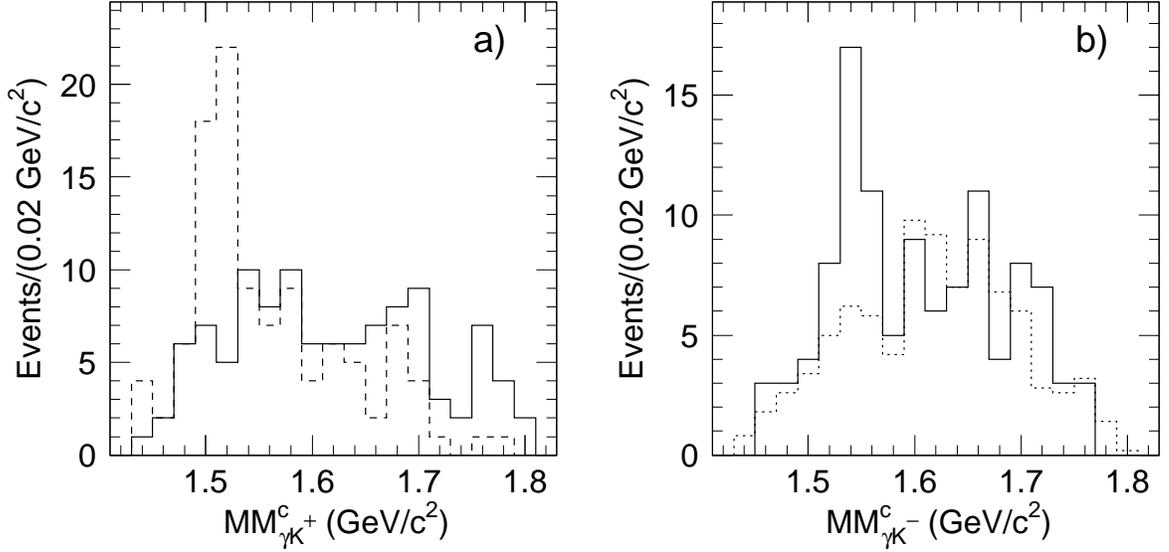}}
\caption{a) The $MM_{\gamma K^+}^c$ spectrum (Eq.~2) for $K^+K^-$
productions for the signal sample (solid histogram) and for events
from the SC with a proton hit in the SSD (dashed histogram).  b) The
$MM_{\gamma K^-}^c$ spectrum for the signal sample (solid histogram)
and for events from the LH$_2$ (dotted histogram) normalized by a fit
in the region above 1.59 GeV/$c^2$.}
\end{figure}

\bibliography{lepsz}

\end{document}